\documentclass[letterpaper, 10 pt, onecolumn, conference]{IEEEtran}  

\IEEEoverridecommandlockouts                              
\usepackage{amssymb, amsfonts, amsmath, fullpage, setspace, graphicx, moreverb, verbatim}

\DeclareMathOperator{\spn}{span}
\DeclareMathOperator{\trace}{trace}
\newcommand{\MM}{\mathcal{M}} 

\newcommand{\Aa}{\mathbf{A}_i}
\newcommand{\Bb}{\mathbf{B}_{i,j}}
\newcommand{\Cc}{\mathbf{C}_i}
\newcommand{\A}{\mathbf{A}}
\newcommand{\B}{\mathbf{B}}
\newcommand{\C}{\mathbf{C}}
\newcommand{\I}{\mathbf{I}}
\newcommand{\R}{\mathbf{R}}
\newcommand{\T}{\mathbf{T}}
\newcommand{\X}{\mathbf{X}_j}
\newcommand{\Y}{\mathbf{Y}_j}
\newcommand{\Z}{\mathbf{Z}}
\newcommand{\0}{\mathbf{0}}
\newcommand{\matrixonetwo}[2]{\left(\begin{array}{cc}#1&#2\\\end{array}\right)}
\newcommand{\matrixonefour}[4]{\left(\begin{array}{cccc}#1&#2&#3&#4\\\end{array}\right)}

\newcommand{\matrixtwoone}[2]{\left(\begin{array}{c}#1\\#2\\\end{array}\right)}
\newcommand{\matrixthreeone}[3]{\left(\begin{array}{c}#1\\#2\\#3\\\end{array}\right)}
\newcommand{\matrixfourone}[4]{\left(\begin{array}{c}#1\\#2\\#3\\#4\\\end{array}\right)}
\newcommand{\matrixsixone}[6]{\left(\begin{array}{c}#1\\#2\\#3\\#4\\#5\\#6\\\end{array}\right)}

\newcommand{\matrixtwotwo}[4]{\left(\begin{array}{cc}#1&#2\\#3&#4\\\end{array}\right)}
\newcommand{\matrixtwothree}[6]{\left(\begin{array}{ccc}#1&#2&#3\\#4&#5&#6\\\end{array}\right)}
\newcommand{\matrixtwofour}[8]{\left(\begin{array}{cccc}#1&#2&#3&#4\\#5&#6&#7&#8\\\end{array}\right)}

\newcommand{\matrixfourtwo}[8]{\left(\begin{array}{cc}#1&#2\\#3&#4\\#5&#6\\#7&#8\\\end{array}\right)}

\begin{document}

\title{Searching for Minimum Storage Regenerating Codes}
\author{\IEEEauthorblockN{Daniel Cullina}
\IEEEauthorblockA{California Institute of Technology}
\and
\IEEEauthorblockN{Alexandros G. Dimakis}
\IEEEauthorblockA{University of Southern California}
\and
\IEEEauthorblockN{Tracey Ho}
\IEEEauthorblockA{California Institute of Technology}
}

\renewcommand{\labelenumi}{\alph{enumi})}

\maketitle
\thispagestyle{empty}
\pagestyle{empty}

\begin{abstract}
Regenerating codes allow distributed storage systems to recover from the loss of a storage node while transmitting the minimum possible amount of data across the network.  We present a systematic computer search for optimal systematic regenerating codes. To search the space of potential codes, we reduce the potential search space in several ways.  We impose an additional symmetry condition on codes that we consider. We specify codes in a simple alternative way, using additional recovered coefficients rather than transmission coefficients and place codes into equivalence classes to avoid redundant checking. Our main finding is a few optimal systematic minimum storage regenerating codes for $n=5$ and $k=3$, over several finite fields. No such codes were previously known and the matching of the information theoretic cut-set bound was an open problem. 
\end{abstract}

\section{Introduction}

Erasure codes can be used in storage systems to efficiently store data while protecting against failures much more efficiently than replication. We can divide a file of size $\MM$ into $k$ pieces, each of size $\MM/k$, encode them into $n$ coded pieces using an $(n,k)$ maximum distance separable (MDS) code, and store them at $n$ nodes. Then, the original file can be recovered from any set of $k$ coded pieces. This is optimal in terms of the redundancy--reliability tradeoff because $k$ pieces, each of size $\MM/k$, provide the minimum data for recovering the file, which is of size $\MM$.

In practical distributed storage systems based on $(n,k)$ MDS codes, we are often faced with the \emph{repair problem}~\cite{DGWR_Infocom07}:
If a node storing a encoded piece fails or leaves the system, in order to maintain the same level of reliability, we need to create a new encoded piece and store it at a new node, but we can only access other encoded blocks. One straightforward way to do so is to let the new node download $k$ encoded pieces from a subset of the surviving nodes, reconstruct the original file, and compute the needed new coded piece. In this process, the new node incurred a total network traffic of $\gamma_{naive}=k\times \MM/k=\MM$.  

Recent prior work~\cite{DGWR_Infocom07} showed that it is possible to reduce this repair bandwidth below $\MM$ and developed information theoretic lower bounds and achievable schemes. At this point, need to distinguish between two different repair problems: 
In this paper we consider the problem of \emph{systematic repair}~\cite{Wu09}
(also called \emph{exact repair}~\cite{Rashmi09}) where we require that it is exactly the same block that is reconstructed after a failure.
This is in sharp contrast to \emph{functional repair} i.e. only requiring that the new block is linearly independent and hence forms a good erasure code jointly with the other existing blocks~\cite{DGWR_Infocom07}. Systematic repair is a strictly harder problem, which however is of great practical interest since in most practical storage systems 
reading parts of the data is the most common operation and it should not require decoding of blocks if no failures have occurred (see also~\cite{ICDCS09} for a practical analysis).

\begin{figure}
\label{fig:53repair}
  \centering
   \vspace{-1cm}
  \includegraphics[width=13cm]{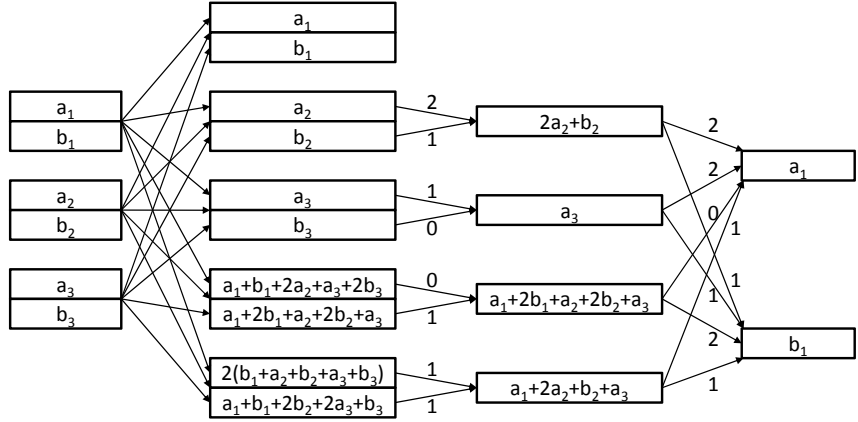}
  \vspace{-3cm}
  \caption{An optimal (5,3) systematic MSR code over GF(3). We show the repair of the first node. The key property is that in the last two packets communicated, the coefficients of $(a_3, b_3)$ are aligned (both are $(1,0)$) and \emph{at the same time}, 
  the coefficients of $(a_2,b_2)$ are aligned (since $(2,1)=2 \times (1,2) \,\text{mod}\, 3$. This allows for only two extra blocks $(2a_2+b_2)$ and $(a_3)$ to suffice for four linear equations that can be solved in the desired variables $(a_1,b_1)$. The rotational symmetry of the code allows all node failures to be recovered similarly}
\end{figure}

As was shown in~\cite{DGWR_Infocom07}, the functional repair problem is equivalent to a multicasting problem on an information flow graph that adds all reconstruction points as virtual data collectors who demand all the data. Using cut-set arguments (which are achievable for multicasting~\cite{Ahlswede00,RLNC}) we can determine the minimum 
repair bandwidth for MDS codes (codes matching this bound are called Minimum Storage Regenerating codes~\cite{DGWR_Infocom07}):
\begin{equation} 
\label{cutset}
\gamma_{MSR}=\frac{\MM}{k} \frac{n-1}{n-k},
\end{equation}
if the new node is allowed to connect to $d=n-1$ surviving nodes, after one failure. 
Note that throughout this paper, we are only considering the minimum storage point and we do not address other points in the storage-repair tradeoff curve of~\cite{DGWR_Infocom07}.

The systematic repair problem, however, is equivalent to a network coding problem where there are receivers who want all the data (the data collectors) and receivers who want subsets of the data (the nodes who will replace the failed ones are now also sinks with a demand of the lost blocks). This reduction shows exactly why systematic repair is a much harder problem and careful coefficient selections are required. 
Further, the cut-set bound~\ref{cutset} is no longer necessarily tight and the optimal systematic repair rates are unknown for general $(n,k)$.  
Recent work~\cite{Wu09} has developed an achievable scheme that is based on aligning the undesired subspaces, similarly to recent ideas for the interference channel (see e.g. 
~\cite{CadambeJ:08}) that have an achievable repair rate of 
\begin{equation}
\label{IR_rate}
\gamma_{IA}=\frac{\MM}{k} \frac{(k-1)(n-k)+1}{n-k},
\end{equation}
achieved by sub-packetizing each packet into $q=n-k$ blocks of size $\MM/kq$ and 
communicating a total of $(k-1)q+1$ blocks from $d=n-1$ surviving nodes. 

It is easy to verify that the achievable rate (\ref{IR_rate}) is matching the cut-set lower bound (\ref{cutset}) for $k=2$ and $k=n-1$ but the other cases remain unknown. 
In this paper we present a searching approach to find systematic MSR codes that match the information theoretic lower bound (\ref{cutset}). Our search found some optimal systematic $(5,3)$ MSR codes (the existence of which was previously unknown), the simplest of which is shown in figure~\ref{fig:53repair}.
The key property that allows optimality is that when one of the undesired subspaces is aligned (as done in the scheme of~\cite{Wu09}), the other is also aligned because of the selection of coefficients of the code. This remarkable property is only possible if the code coefficients are carefully chosen and is closely linked to the size of the finite field. 
  
To search the space of potential codes in feasible amounts of time, we reduce the search space in several ways.  We impose an additional condition that restricts the type of codes that we consider.  This allows us to consider only highly symmetric codes that can be more concisely specified.  We specify a code in a simple alternative way, using additional recovered coefficients rather than transmission coefficients.  The space of codes can be searched more easily and efficiently when codes are specified this way.  Finally, we use linear transformations to relate codes to each other and place them into equivalence classes.  This allows us to check only one code from each equivalence class.

\section{Definitions and Notation}
The storage networks that we are concerned with contain $n$ equivalent storage nodes.  We wish to store $\mathcal{M}$ bits of data in the network, where $\mathcal{M}$ is	$k$ times the size of one of the storage nodes.  Because of this, we say that the network has $k$ source nodes.

\subsection{Lower bound on recovery bandwidth}

During the recovery process, $\frac{\mathcal{M}d}{k(d-k+1)}$ bits of data must be transmitted, where $d$ is the number of nodes providing data \cite{DGWR_Infocom07}.  We are interested in the case where $d=n-1$, so this bound becomes $\frac{\mathcal{M}(n-1)}{k(n-k)}$.  There are $n-1$ nodes that each contain $\frac{\mathcal{M}}{k}$, so each node is transmitting $\frac{1}{n-k}$ of its contents.  Because of this, we store $n-k$ packets of data in each storage node.  We break the source data up into packets of the same size and each storage packet will be some linear combination of the $k(n-k)$ packets of source data.  

\subsection{Notation} 
	We use several matrices to represent the data and coefficients used in an MSR code.
	\begin{center}
	\begin{tabular}{|l|ccc|l|}
	\hline
	  $\Aa$ & $(n-k)$ & $\times$ & $k(n-k)$ & matrix of storage coefficients\\
	  \hline
  	$\Bb$ & $1$ & $\times$ & $(n-k)$ & row vector of transmission coefficients\\
  	\hline
  	$\Cc$ & $(n-k)$ & $\times$ & $(n-1)$ & matrix used to rebuild storage node $i$\\
  	\hline
  	$\mathbf{D}$ & $k(n-k)$ & $\times$ & $x$ & matrix of source data\\
  	\hline
	\end{tabular}
	\end{center}
	The $i$th storage node contains $\Aa\mathbf{D}$, the original data multiplied by the storage coefficients for that node.

\subsection{Independence}
		The storage nodes of the code are independent if any $k$ nodes can reproduce the original data.  That is, for all combinations of $k$ storage nodes, there is a matrix $\mathbf{M}$ such that\\
	\begin{eqnarray}
	\mathbf{D} &=& 
	\mathbf{M}
	\matrixfourone
		{\A_{c(1)}}
		{\A_{c(2)}}
		{\vdots}
		{\A_{c(k)}}
	\mathbf{D}
	\end{eqnarray}
	for any value of D.  An equivalent condition is that for each combination of k nodes, the matrix of storage coefficients must have full rank, i.e.~a nonzero determinant.

\subsection{Recovery}
	When node $j$ fails, the $i$th node transmits $\Bb\Aa\mathbf{D}$.	The code allows the recovery of node $j$ if there is a matrix $\C_j$ that recreates the lost node from the transmitted vectors:
	\begin{eqnarray}
	\A_{j}\mathbf{D} &=& 
	\C_j
	\matrixsixone
    {\B_{1,j}\A_{1}}
    {\vdots}
    {\B_{j-1,j}\A_{j-1}}
    {\B_{j+1,j}\A_{j+1}}
    {\vdots}
		{\B_{n,j}\A_{n}}
	\mathbf{D}
	\end{eqnarray}
	 for any value of $\mathbf{D}$.  Therefore, $\mathbf{D}$ drops out of both the independent and recovery conditions, and we can focus on the coding coefficients only.  We can also ignore the $\C_i$ matrices because from the recovery condition we can see that in a working code the $\Cc$ matrices are fully specified by the $\Aa$ and $\Bb$ matrices.  With these two conditions, we can determine whether a set of $\Aa$ and $\Bb$ matrices form a code.   
\section{Rotationally Symmetric Codes}

	To reduce the total number of coefficients, we consider codes whose $\Aa$ matrices are related to each other by a simple transformation.\\
	Let $\R$ be an $k(n-k)\times k(n-k)$ matrix such that 
	\begin{eqnarray}
	\R^{n}=\I,
	\end{eqnarray}
	and let 
	\begin{eqnarray}
	\Aa=\A\R^{i}.
	\end{eqnarray}\\
	A discussion of the $\R$ matrices themselves can be found in~\cite{DanThesis}.
	This reduces the number of storage coefficients needed to specify a code by a factor of $n$, reducing the search space exponentially.
	
\subsection{Recovery Condition}
	This makes the recovery condition\\
	\begin{eqnarray}
	\A\R^{j} &=& 
	  \C_{j}
	  \matrixsixone
	    {\B_{1,j}\A\R^{1}}
	    {\vdots}
	    {\B_{j-1,j}\A\R^{j-1}}
	    {\B_{j+1,j}\A\R^{j+1}}
	    {\vdots}
			{\B_{n,j}\A\R^{n}}
			\\
		\A &=& 
	  \C_{j}
	  \matrixsixone
	    {\B_{1,j}\A\R^{n-j+1}}
	    {\vdots}
	    {\B_{j-1,j}\A\R^{n-1}}
	    {\B_{j+1,j}\A\R^{1}}
	    {\vdots}
			{\B_{n,j}\A\R^{n-j}}.
		\end{eqnarray}
We can replace $\Bb$ with $\B_{i-j}$, reorder the rows of the transmitted coefficient matrix, and replace $\C_{j}$ with  $\C$.  Now there is only one recovery condition.\\
		\begin{eqnarray}
		\A &=&
	  \C
	  \matrixthreeone
	    {\B_{1}\A\R^{1}}
	    {\vdots}
			{\B_{n-1}\A\R^{n-1}}
		\end{eqnarray}
		This is an improvement of a factor of $n$.
\subsection{Independence Condition}
	Similarly, when checking independence, we only need to check combinations that include the first node.  
	\begin{eqnarray}
	  \det\matrixfourone
			{\A\R^{c(1)}}
			{\A\R^{c(2)}}
			{\vdots}
			{\A\R^{c(k)}}
		&=&
		\det\matrixfourone
			{\A\R^1}
			{\A\R^{c(2)-c(1)+1}}
			{\vdots}
			{\A\R^{c(k)-c(1)+1}}
		\det\R^{c(1)}
		\end{eqnarray}
	
	This reduces the number of conditions from $\matrixtwoone{n}{k}$ to $\matrixtwoone{n-1}{k-1}$.  This is an improvement of a factor of $\frac{n}{k}$.\\

\subsection{Example}
	\begin{eqnarray}
		\R &=& \left(\begin{array}{cccc}0&1&0&0\\0&0&1&0\\0&0&0&1\\1&0&0&0\\\end{array}\right)\\
		\A_1 &=& \matrixtwofour{1}{0}{0}{0} {0}{1}{1}{0}
	\end{eqnarray}
	The $\B_{i}$ matrices gives us the transmitted vectors.
	\begin{eqnarray}
		\B_{1}\A_2 &=& 
		\matrixonetwo{1}{0}\matrixtwofour{0}{1}{0}{0} {0}{0}{1}{1}\\
		&=& \matrixonefour{0}{1}{0}{0}
	\end{eqnarray}
	\begin{eqnarray}
		\B_{2}\A_3 &=& 
		\matrixonetwo{1}{0}\matrixtwofour{0}{0}{1}{0} {1}{0}{0}{1}\\
		&=& \matrixonefour{0}{0}{1}{0}
	\end{eqnarray}
	\begin{eqnarray}
		\B_{3}\A_4 &=& 
		\matrixonetwo{0}{1}\matrixtwofour{0}{0}{0}{1} {1}{1}{0}{0}\\
		&=& \matrixonefour{1}{1}{0}{0}
	\end{eqnarray}
	From these we can complete the code by calculating $\C$.
	\begin{eqnarray}
		\A_1 &=& \C
			\matrixthreeone 
				{\B_{1}\A_2}
				{\B_{2}\A_3}
				{\B_{3}\A_4}
			\\
			\matrixtwofour{1}{0}{0}{0} {0}{1}{1}{0} &=& 
			\C\left(\begin{array}{cccc}0&1&0&0\\0&0&1&0\\1&1&0&0\\\end{array}\right)
			\\
			\C &=& \matrixtwothree{-1}{0}{1} {1}{1}{0}
	\end{eqnarray}
		
\section{Additional Recovered Coefficients}

	  The $\Bb$ matrices cannot be eliminated in a similarly simple manner, but their contribution to the code to be represented in alternative way.  During recovery $n-1$ vectors are transmitted to the lost node, but the original $\Aa$ matrix has only $n-k$ rows.  Thus $k-1$ additional vectors of coefficients are recovered. Specifying these vectors allows the $\Bb$ matrices to be determined.
	  
	Let $\Z_j$ be the $k-1 \times k(n-k)$ matrix that contains the additional rows recovered when node $j$ is lost.
Let 
\begin{eqnarray}
\X=\matrixtwoone{\Z_j}{\A_j}
\end{eqnarray}
be the $n-1 \times k(n-k)$ matrix that contains all of the rows recovered when node $j$ is lost. Then $\X^{T}(\X\X^{T})^{-1}\X$ projects vectors into $\spn \X$. A row vector $\mathbf{v}$ is in $\spn{\X}$ if the projection does not change the vector, or 
\begin{eqnarray}
\mathbf{v}(\X^{T}(\X\X^{T})^{-1}\X)=\mathbf{v}
.
\end{eqnarray}
This can be rewritten as 
\begin{eqnarray}
\mathbf{v}(\I-\X^{T}(\X\X^{T})^{-1}\X)=\mathbf{0}.
\end{eqnarray}
	
	$\I-\X^{T}(\X\X^{T})^{-1}\X$ gives the difference between the original vector and the projection.	This is a projection to the $(k-1)(n-k-1)$-dimensional space $\mathbb{F}^n/\spn{\X}$.  
The only potentially useful vectors to transmit during recovery are those in $\spn \X$, so we need to ensure that the transmitted vector $\Bb\Aa$ must satisfy 
\begin{eqnarray}
\Bb\Aa(\I-\X^{T}(\X\X^{T})^{-1}\X)=\mathbf{0}.  
\end{eqnarray}
Thus the choices for $\Bb$ are the vectors in the nullspace of $\Aa(\I-\X^{T}(\X\X^{T})^{-1}\X)$.

\subsection{Unrecovered Coefficients}
Let $\Y$ refer to a basis that spans $\mathbb{F}^n/\spn{\X}$.  Now we can rewrite the projection as $\I-\X^{T}(\X\X^{T})^{-1}\X=\Y^{T}(\Y\Y^{T})^{-1}\Y$.  Now we can say $\Bb\Aa$ must satisfy  $\Bb\Aa\Y^{T}(\Y\Y^{T})^{-1}\Y=\mathbf{0}$, which reduces to $\Bb\Aa\Y^{T}=0$.  Thus the null space of $\Aa(\I-\X^{T}(\X\X^{T})^{-1}\X)$ is the same as the nullspace of $\Aa\Y^{T}$.  $\Aa\Y^{T}$ is a $(n-k)\times(k-1)(n-k-1)$ matrix, so its nullity is at least $(n-k)-(k-1)(n-k-1)$ or $1+(n-k-1)(2-k)$.  However, if $2<k<n-1$, this bound does not force the nullity to be positive.  This bound does explain why it is so easy to find codes when $k=2$.

\subsection{Example: Obtaining $\B$ from $\mathbf{Y}$}
	Now we can see how the $\Bb$ vectors were discovered in the previous example.  Let	$\mathbf{Y}_1 = (\begin{array}{cccc} 0&0&0&1\\ \end{array})$.  Note that $\A_1\mathbf{Y}_1^T=0$ as required.  We apply $\mathbf{Y}_1^T$ to the other $\Aa$ matrices and find the $\Bb$ vectors that satisfy $\Bb\Aa\Y^{T}=0$.
	\begin{eqnarray}
	\A_2\mathbf{Y}_1^T &=& \matrixtwoone{0}{1} \\ 
	\B_{1} &=& \matrixonetwo{1}{0}
	\end{eqnarray}
	\begin{eqnarray}
	\A_3\mathbf{Y}_1^T &=& \matrixtwoone{0}{1} \\ 
	\B_{2} &=& \matrixonetwo{1}{0}
	\end{eqnarray}
	\begin{eqnarray}
	\A_4\mathbf{Y}_1^T &=& \matrixtwoone{1}{0} \\ 
	\B_{3} &=& \matrixonetwo{0}{1}
	\end{eqnarray}
	For the $n=4$, $k=2$ case, nearly all choices for $\Y$ produce a working code.  This is not the case for larger coefficients.  
		 
\section{Transformations of codes and equivalence classes}
\subsection{Row transformations}
Suppose we have an invertible $(n-k)\times(n-k)$ matrix $\T$ and a working code defined by $\Aa$ and $\Bb$ matrices.  Then the matrices $\T\Aa$ and $\Bb\T^{-1}$ also define a working code. For recoverability we have
	\begin{eqnarray}
	\T\A_{j} &=& 
	  \T\C_{j}
	  \matrixsixone
	    {\B_{1,j}\T^{-1}\T\A_{1}}
	    {\vdots}
	    {\B_{j-1,j}\T^{-1}\T\A_{j-1}}
	    {\B_{j+1,j}\T^{-1}\T\A_{j+1}}
	    {\vdots}
			{\B_{n,j}\T^{-1}\T\A_{n}}
		\\
		&=&\T\C_{j}
	  \matrixsixone
	    {\B_{1,j}\A_{1}}
	    {\vdots}
	    {\B_{j-1,j}\A_{j-1}}
	    {\B_{j+1,j}\A_{j+1}}
	    {\vdots}
			{\B_{n,j}\A_{n}},
		\end{eqnarray}
	and for independence we have
	\begin{eqnarray}
	  \det\matrixfourone
			{\T\A_{c(1)}}
			{\T\A_{c(2)}}
			{\vdots}
			{\T\A_{c(k)}}
		&=&
		\det\left(\begin{array}{cccc}
			\T & 0 & \hdots & 0\\
			0 & \T & \ddots & \vdots\\
			\vdots & \ddots & \ddots & 0\\
			0 & \hdots & 0 & \T\\
		\end{array} \right)
		\det\matrixfourone
			{\A_{c(1)}}
			{\A_{c(2)}}
			{\vdots}
			{\A_{c(k)}}
		\\
		&=&(\det\T)^{k}\det\matrixfourone
			{\A_{c(1)}}
			{\A_{c(2)}}
			{\vdots}
			{\A_{c(k)}}
		\\
		&\neq&0.
		\end{eqnarray}
		The row transformation is applied to the $\A$ matrices from the left and the rotation matrix in a rotationally symmetric code is applied from the right.  Thus, applying the transformation to a rotationally symmetric code results in another rotationally symmetric code that uses the same rotation matrix.  We can define codes to be equivalent if they are related by a row transformation.  Testing only one code from each equivalence class reduces the search space by $k^2$ dimensions.
		
\subsection{Column transformations}
The same technique can be applied to the columns.  If we have an invertible $k(n-k)\times k(n-k)$ matrix $\T$ and a working code defined by $\Aa$ and $\Bb$ matrices, then the matrices $\Aa\T$ and $\Bb$ also define a working code. For recoverability we have
	\begin{eqnarray}
	\A_{j}\T &=& 
	  \C_{j}
	  \matrixsixone
	    {\B_{1,j}\A_{1}\T}
	    {\vdots}
	    {\B_{j-1,j}\A_{j-1}\T}
	    {\B_{j+1,j}\A_{j+1}\T}
	    {\vdots}
			{\B_{n,j}\A_{n}\T}
		\\
		&=&\C_{j}
	  \matrixsixone
	    {\B_{1,j}\A_{1}}
	    {\vdots}
	    {\B_{j-1,j}\A_{j-1}}
	    {\B_{j+1,j}\A_{j+1}}
	    {\vdots}
			{\B_{n,j}\A_{n}}
		\T,
	\end{eqnarray}
	and for independence we have
	\begin{eqnarray}\det\matrixfourone
			{\A_{c(1)}\T}
			{\A_{c(2)}\T}
			{\vdots}
			{\A_{c(k)}\T}
		&=& \det\matrixfourone
			{\A_{c(1)}}
			{\A_{c(2)}}
			{\vdots}
			{\A_{c(k)}}
		\det\T
		\\
		&\neq&0.
		\end{eqnarray}
		In a rotationally symmetric code, the column transformation and the rotation are both applied from the right, so they interact.
		\begin{eqnarray}
		\Aa\T &=& \A\R^{i}\T\\
		&=& \A\T\T^{-1}\R^{i}\T\\
		&=&	\A\T(\T^{-1}\R\T)^{i} 
		\end{eqnarray}
		So the new code is rotationally symmetric with a different rotation matrix, $\T^{-1}\R\T$.  This means that we can use a simple rotation matrix when searching for codes and simultaneously check all rotationally symmetric codes that use similar rotation matrices.
		
		This also makes it possible to put any rotationally symmetric code into systematic form.  When a code is in systematic form, the first $k$ storage matrices can be stacked to form an identity matrix. 
\begin{eqnarray}
			\matrixthreeone
			{\A_1\T}
			{\vdots}
			{\A_k\T}
		&=& \I
\end{eqnarray}
Finding the transformation that puts a code into systematic form is simple.  It is simply the inverse of the stack of first $k$ storage matrices.

\begin{eqnarray}
	\T &=&
	\matrixthreeone
		{\A_1}
		{\vdots}
		{\A_k}
	^{-1}
\end{eqnarray}
\subsection{Example: Systematic Form}
\begin{eqnarray}
\matrixtwoone{\A_1}{\A_2} &=& 
	\left(\begin{array}{cccc}
		1&0&0&0\\
		0&1&1&0\\
		0&1&0&0\\
		0&0&1&1\\
	\end{array} \right)
\\
\T &=& \matrixtwoone{\A_1}{\A_2}^{-1}
\\
&=&
	\left(\begin{array}{cccc}
		1&0&0&0\\
		0&0&1&0\\
		0&1&-1&0\\
		0&-1&1&1\\
	\end{array} \right)
\end{eqnarray}

The same $\B$ vectors as before will work for recovery.
\begin{eqnarray}	
	\B_{1}\A_2\T &=& 
	\matrixonetwo{1}{0}\matrixtwofour{0}{0}{1}{0} {0}{0}{0}{1}\\
	&=& \matrixonefour{0}{0}{1}{0}
\end{eqnarray}
\begin{eqnarray}
	\B_{2}\A_3\T &=& 
	\matrixonetwo{1}{0}\matrixtwofour{0}{1}{-1}{0} {1}{-1}{1}{1}\\
	&=& \matrixonefour{0}{1}{-1}{0}
\end{eqnarray}
\begin{eqnarray}
	\B_{3}\A_4\T &=& 
	\matrixonetwo{0}{1}\matrixtwofour{0}{-1}{1}{1} {1}{0}{1}{0}\\
	&=& \matrixonefour{1}{0}{1}{0}
\end{eqnarray}

The same $\C$ matrix as before will also work.
\begin{eqnarray}
	\A_1\T &=& \C
		\matrixthreeone 
			{\B_{1}\A_2}
			{\B_{2}\A_3}
			{\B_{3}\A_4}
			\T
	\\
	&=& \matrixtwothree{-1}{0}{1} {1}{1}{0}
	\left(\begin{array}{cccc}0&0&1&0\\0&1&-1&0\\1&0&1&0\\\end{array}\right)
	\\
	&=& \matrixtwofour{1}{0}{0}{0} {0}{1}{0}{0} 
\end{eqnarray}

\section{Search Procedure}

When searching for codes of a given $n$ and $k$ over a finite field, this procedure was used.  Iterate over $\A$ matrices in a way that ensures that exactly one matrix from each row transformation equivalence class is produced.  For each $\A$ matrix, produce the collection of $n$ $\Aa$ matrices using a single simple column rotation matrix.  Then test the independence condition.  Test it before the recovery condition because it requires only $\Aa$ matrices.  If the independence condition is met, iterate over the space of potential additional recovered coefficients.  For each $\X$ matrix produced by this process, check the recovery condition.  If the condition is met, this is a code.

\section{Search results}

\subsection{$n=4$, $k=2$}

These coefficients are small enough to all several fields to be searched exhaustively.  We have searched the prime fields up to $GF(13)$.  There are no rotationally symmetric codes in $GF(2)$, but in all larger fields codes are extremely easy to find.  In all of these fields, nearly all of the potential codes that satisfy the independence condition also satisfy the recovery condition.  As the field size increases, larger and larger fractions of the potential codes satisfy the independence condition.
In $GF(3)$, $22\%$ of potential codes satisfy the independence condition, and of these all satisfy the recovery condition.  In $GF(13)$, $78\%$ of potential codes satisfy the independence condition and of these $92\%$ also satisfy the recovery condition.

\subsection{$n=5$, $k=3$}

For these coefficients, codes were not previously known.  We have exhaustively searched $GF(2)$, $GF(3)$, $GF(4)$, and $GF(5)$ and randomly searched in larger fields for rotationally symmetric codes.  We found codes in $GF(3)$, $GF(4)$, $GF(7)$, and larger fields, but none in $GF(2)$ or $GF(5)$.  While the codes we have found in smaller fields are not composed of vectors in general position, we found a code in $GF(17)$ that is.  Several of these codes are given in the appendix.  The full descriptions can be found in~\cite{DanThesis}.

\subsection{$n=6$, $k=3$}

For these coefficients, We have yet to find any codes.  In $GF(3)$, only about $1\%$ of potential codes satisfy the independence condition.  In $GF(4)$ this number is about $14\%$ and in $GF(5)$ it is about $30\%$.

\bibliography{thesisbib}{}
\bibliographystyle{IEEEtran}

\section{Appendix}

\subsection{$(5,3)$ code over $GF(3)$ in systematic form}
\begin{eqnarray}
	\B_{2,1}\A_2 &=& 
	\left(\begin{array}{cc} 2&1\\ \end{array}\right)
	\left(\begin{array}{cccccc} 0&0&1&0&0&0\\0&0&0&1&0&0\\ \end{array}\right)
\end{eqnarray}
\begin{eqnarray}
	\B_{3,1}\A_3 &=& 
	\left(\begin{array}{cc} 1&0\\ \end{array}\right)
	\left(\begin{array}{cccccc} 0&0&0&0&1&0\\0&0&0&0&0&1\\ \end{array}\right)
\end{eqnarray}
\begin{eqnarray}
	\B_{4,1}\A_4 &=& 
	\left(\begin{array}{cc} 0&1\\ \end{array}\right)
	\left(\begin{array}{cccccc} 1&1&2&0&1&2\\1&2&1&2&1&0\\ \end{array}\right)
\end{eqnarray}
\begin{eqnarray}
	\B_{5,1}\A_5 &=& 
	\left(\begin{array}{cc} 1&1\\ \end{array}\right)
	\left(\begin{array}{cccccc} 0&2&2&2&2&2\\1&1&0&2&2&1\\ \end{array}\right)
\end{eqnarray}
\begin{eqnarray}
	\A_1 &=& \C
		\left(\begin{array}{c} 
			\B_{2,1}\A_2\\
			\B_{3,1}\A_3\\
			\B_{4,1}\A_4\\
			\B_{5,1}\A_5\\
		\end{array}\right)
	\\
		&=& 
		\left(\begin{array}{cccc} 2&2&0&1\\1&0&2&1\\ \end{array}\right)
		\left(\begin{array}{cccccc} 0&0&2&1&0&0\\0&0&0&0&1&0\\1&2&1&2&1&0\\1&0&2&1&1&0\\ \end{array}\right)
\end{eqnarray}

\subsection{$(5,3)$ code over $GF(7)$ in systematic form}
\begin{eqnarray}
	\B_{2,1}\A_2 &=& 
	\left(\begin{array}{cc} 0&1\\ \end{array}\right)
	\left(\begin{array}{cccccc} 0&0&1&0&0&0\\0&0&0&1&0&0\\ \end{array}\right)
\end{eqnarray}
\begin{eqnarray}
	\B_{3,1}\A_3 &=& 
	\left(\begin{array}{cc} 2&1\\ \end{array}\right)
	\left(\begin{array}{cccccc} 0&0&0&0&1&0\\0&0&0&0&0&1\\ \end{array}\right)
\end{eqnarray}
\begin{eqnarray}
	\B_{4,1}\A_4 &=& 
	\left(\begin{array}{cc} 5&1\\ \end{array}\right)
	\left(\begin{array}{cccccc} 2&0&5&6&1&1\\6&4&3&4&5&0\\ \end{array}\right)
\end{eqnarray}
\begin{eqnarray}
	\B_{5,1}\A_5 &=& 
	\left(\begin{array}{cc} 6&1\\ \end{array}\right)
	\left(\begin{array}{cccccc} 1&4&3&3&4&0\\3&0&3&6&1&2\\ \end{array}\right)
\end{eqnarray}
\begin{eqnarray}
	\A_1 &=& \C_1
		\left(\begin{array}{c} 
			\B_{2,1}\A_2\\
			\B_{3,1}\A_3\\
			\B_{4,1}\A_4\\
			\B_{5,1}\A_5\\
		\end{array}\right)
		\\
		&=& 
		\left(\begin{array}{cccc} 3&0&2&2\\4&4&1&6\\ \end{array}\right)
		\left(\begin{array}{cccccc} 0&0&0&1&0&0\\0&0&0&0&2&1\\2&4&0&6&3&5\\2&3&0&3&4&2\\ \end{array}\right)	
\end{eqnarray}

\end{document}